\documentclass[aps,pra,twocolumn,showpacs,amsmath,amssymb]{revtex4-1}

\usepackage{graphicx}
\usepackage{color}
\usepackage{epstopdf}
\usepackage{amsmath}
\usepackage{bm}
\usepackage{hyperref}
\usepackage{cleveref}
\usepackage{tcolorbox}
\usepackage[export]{adjustbox}

\newcommand{\reffig}[1]{Fig.~\ref{#1}}
\newcommand{\refeq}[1]{Eq.~(\ref{#1})}

\graphicspath{{./figures/}}

\begin{document}

\title{Metastable Patterns in one- and two-component dipolar Bose-Einstein Condensates}

\author{Yong-Chang Zhang$^{1}$}
\email{zhangyc@xjtu.edu.cn}
\author{Thomas Pohl$^{2}$}
\author{Fabian Maucher$^{3,4}$}
\affiliation{$^1$MOE Key Laboratory for Nonequilibrium Synthesis and Modulation of Condensed Matter, Shaanxi Key Laboratory of Quantum Information and Quantum Optoelectronic Devices, School of Physics, Xi’an Jiaotong University, Xi’an 710049, People’s Republic of China \\ $^2$Department of Physics and Astronomy, Aarhus University, Ny Munkegade 120, DK 8000 Aarhus, Denmark \\ 
$^3$Faculty of Mechanical, Maritime and Materials Engineering; Department of Precision and Microsystems Engineering, Delft
University of Technology, 2628 CD, Delft, The Netherlands\\
$^4$Departament de Física, Universitat de les Illes Balears \& IAC-3, Campus UIB, E-07122 Palma de Mallorca, Spain}

\begin{abstract}
 In this paper we study metastable states in single- and two-component dipolar Bose-Einstein condensates. We show that this system supports a rich spectrum of symmetries that are remarkably stable despite not being ground states. In a parameter region where striped phases are ground states, we find such metastable states that are energetically favourable compared to triangular and honeycomb lattices. Among these metastable states we report a peculiar ring-lattice state, which is led by the competition between triangular and honeycomb symmetries and rarely seen in other systems. In the case of dipolar mixtures we show that via tuning the miscibility these states can be stabilized in a broader domain by utilising inter-species interactions. 
\end{abstract}
\maketitle

\section{Introduction}

Ultracold quantum gases with long-range interactions display remarkably intriguing behaviour and give access to probing fundamental quantum behaviour~\cite{Bloch:RMP:2008,Mollmer:RMP:2010}. 
Dipolar Bose-Einstein condensates (BECs) permit a controlled access to such effects~\cite{Pfau:RepProgPhys:2009,Chomaz_2023_review}. 
These include the recent experimental advances in the observation of quantum droplets~\cite{Pfau:nature2:2016,Pfau:PRL:2016,Pfau:PRR:2019}, supersolids
~\cite{Pfau:nature:2016,Ferlaino:PRX:2016,Ferlaino:PRX:2019,Modugno:PRL:2019,Pfau:PRX:2019,Pfau:PRX:2021,Tanzi:Science:2021,Bisset:PRL:2022,Ferlaino:PRL:2022} and their excitation spectra~\cite{Lewenstein:PRL:2003,Guo:Nature:2019,Tanzi:Nature:2019,Natale:PRL:2019}. The existence of these states of matter can be viewed as a macroscopic signature for quantum fluctuations in that they suppress dipolar collapse that would otherwise occur~\cite{Lahaye:PRL:2008}. 
With that dipolar quantum gases emerged as an ideal platform to observe fundamentally interesting and surprising  physical effects. 
One of these effects is the emergence of a point in phase-space where the superfluid-supersolid phase-transition becomes second-order~\cite{Yongchang:PRL:2019},  which means around that point the physics becomes practically linear in the modulation amplitude and acquires a glass-like nature and rich variety of patterns~\cite{Pfau:PRR:2021,Yongchang:PRA:2021,Blakie:arxiv:2023}. 

Generally, considering more than one species of atoms in ultracold quantum gases~\cite{Cornish:PRA:2011,Arlt:PRA:2015} adds complexity and 
can promote a range of intricate phenomena, including collapse suppression even for short-ranged interactions~\cite{Petrov:PRL:2015,Tarruell:Science:2018} by quantum-fluctuations in mixtures as well as tunable miscibility~\cite{Cornell:PRL:1998,Wiemann:PRL:2008} between the components. 

The recent observation of dipolar mixtures~\cite{Ferlaino:PRL:2018,Ferlaino:PRA:2020,Ferlaino:PRA:2022} paves the way for a range of new perspectives and qualitatively new behaviour as it permits the combination of the intriguing behaviour of two-component physics with long-range interactions~\cite{Santos:PRL:2021,Blakie:PRL:2021,Bisset:PRA:2021,Santos:PRA:2023}. 
 
Metastability is ubiquitous in nature ranging from physics~\cite{Kosterlitz:JPC:1973,Apaja:EPL:2008}, chemistry~\cite{Brazhkin:PhyUsp:2006} to material~\cite{Yoshida:NANO:2018} science. Metastabilty can lead to a variety of rich phenomena~\cite{Ueda:PRL:2008,Yi:PRL:2007,Lewenstein:PRL:2007,Lewenstein:PRA:2008,Carlos:PRL:2009,Blass:PRL:2018} and permits gaining further insights into the overall physics of the systems. 
The transition between metastable states has attracted attention as well~\cite{Hruby:PNAS:2018}. Previous work have found that long-range interaction is a crucial ingredient for the appearance of metastability \cite{Yi:PRL:2007,Lewenstein:PRL:2007,Lewenstein:PRA:2008,Carlos:PRL:2009,Blass:PRL:2018,PhysRevB.105.094505,RevModPhys.95.035002}. For example, it has been reported that dipolar atoms loaded in optical lattices can host a number of metastable states in Mott-insulator regimes~\cite{Lewenstein:PRL:2007,Lewenstein:PRA:2008}. However, it is still unclear whether a continuous diploar gas can support metastable phases that feature a significantly different symmetry to the groundstate as well. 

Here, we explore such metastable states that feature multiple length scales in single- as well as two-component dipolar BECs. 
Close to the second-order point deformations of small density modulations barely lead to a change in energy due to the shallowness of the energy-landscape. 
Therefore, one can imagine that states with different symmetry can be created via linear superposition, however, such superpositions can be expected to be unstable. However, the dynamics can become so slow that the density almost appears frozen. 
Further away from the second-order point, where the periodic density modulations become larger and interactions lead to a more pronounced energy landscape, the possibility of sufficiently deep local energy minima appears more reasonable. 

To address that idea systematically, we start off with considering single-component systems and investigate the emergence of superlattices by superposing two different patterns in regions close to lines where they are energetically degenerate. 
To avoid finite-size effects we consider the thermodynamic limit. By thermodynamic limit we refer to the situation where the plane perpendicular to the dipolar polarization direction $z$, which corresponds to both trapping and polarisation axis, is infinitely extended. In this plane, the average two-dimensional (2D) density $\rho_{\rm 2D}$ is fixed, and a 2D symmetry breaking occurs. 

After studying the single-component system we consider dipolar mixtures. Here, we have the additional degrees of freedom due to the interaction between the two components. 
By changing the miscibility we can tune between a situation where both components reach their maximum density at the center of the trap and triple-layered density distributions, where one component is ``sandwiched" between two layers of the other component. The interaction between the layers might render such metastable states unstable or possibly stabilize them. 

\section{Modelling of Dipolar BECs}

\subsection{Single-component BECs} 
A single-component dipolar Bose-Einstein condensate at zero-temperature composed of  $N$ dipolar Bosonic atoms of mass $m$ including quantum fluctuations can be described via 
\begin{align}
i\frac{\partial}{\partial t}&\psi({\bf r}) = \Big [-\frac{\nabla^2}{2} + \frac{1}{2}\omega^2_z z^2 +  g \rho({\bf r}) \nonumber \\
&+ \int {\rm d}^3 {\bf r}' V({\bf r}-{\bf r}')\rho({\bf r}')  + \mu_{\rm LHY} \Big ] \psi({\bf r}), 
\label{eq:NLS}
\end{align}
Here, $\rho({\bf r})\equiv|\psi({\bf r})|^2$ represents the condensate density, we assume trapping along the polarization direction only, in this case the $z$-direction, and $\omega_z$ is the respective frequency of the harmonic trap. 
Furthermore, $g=\frac{a_{\rm s}}{3a_{\rm dd}}$ is the fraction of the s-wave scattering length $a_s$ to the dipolar length $a_{\rm dd}$, the latter of which characterizing the dipole-dipole interaction strength. $V({\bf r})=\frac{1}{4\pi r^3}(1-3z^2/r^2)$ is the usual dipole-dipole interaction. The Lee-Huang-Yang (LHY) correction due to quantum fluctuations $\mu_{\rm LHY}$ is given by~\cite{Pelster:PRA:2011, Pelster:PRA:2012, Blakie:PRA:2016, Santos:PRA:2016, Bisset:PRA:2016}. 
\begin{align}
    \mu_{\rm LHY}=\frac{4}{3\pi^2}(\frac{a_{\rm s}}{3a_{\rm dd}})^{5/2}[1+\frac{3}{2}(\frac{a_{\rm dd}}{a_{\rm s}})^2]\rho^{3/2}
\end{align}

\subsection{Two-component BECs}
To generalize the single-component description to model two-component dipolar condensates we assume the local density approximation and employ the model recently introduced in \cite{Santos:PRL:2021,Blakie:PRL:2021}:
\begin{align}
i\frac{\partial}{\partial t}&\psi_\alpha({\bf r}) = \Big [-\frac{\nabla^2}{2} + \frac{1}{2}\omega^2_z z^2 + \sum_{\beta} \frac{a^{\alpha\beta}_{\rm s}}{3a^{11}_{\rm dd}} \rho_{\beta}({\bf r}) \nonumber \\
&+ \sum_{\beta}\int {\rm d}^3 {\bf r}' V_{\alpha \beta}({\bf r}-{\bf r}')\rho_{\beta}({\bf r}')  + \mu_{\rm LHY}^{(\alpha)} \Big ] \psi_\alpha({\bf r}), 
\label{eq:NLS}
\end{align}
where $\alpha,\beta=1,2$, $\rho_\alpha({\bf r})\equiv|\psi_\alpha({\bf r})|^2$, $\omega_z$ is the frequency of the harmonic trap along the polarization direction, $V_{\alpha\beta}({\bf r})=\frac{\sqrt{a^{\alpha \alpha}_{\rm dd} a^{\beta \beta}_{\rm dd}}}{a^{11}_{\rm dd}}\frac{1}{4\pi r^3}(1-3z^2/r^2)$ is the usual dipole-dipole interaction with $a^{\alpha\beta}_{\rm s}$ being the s-wave scattering length and $a^{\alpha\alpha}_{\rm dd}$ the typical dipolar length characterizing the dipole-dipole interaction strength. And the LHY correction $\mu_{\rm LHY}^{(\alpha)}$ is given by 
\begin{equation}
     \mu_{\rm LHY}^{(\alpha)}=\frac{1}{3\sqrt{2}\pi^2}\int^{\pi/2}_0 d\theta \sin{\theta} \left( \mathcal{I}^+_\alpha + \mathcal{I}^-_\alpha \right)
\end{equation}
with $\mathcal{I}^\pm_\alpha=\left(u_{\alpha} \pm \frac{(-1)^{\alpha-1}\delta \rho_{\alpha} + 2 u^2_{12} \rho_{3-\alpha}}{\sqrt{\delta^2+4u^2_{12}\rho_1 \rho_2}} \right) \mathcal{J}^{3/2}$, $\mathcal{J}= u_{11} \rho_1 + u_{22} \rho_2 \pm \sqrt{\delta^2+4u^2_{12}\rho_1 \rho_2} $, $u_{\alpha\beta}=\frac{a^{\alpha\beta}_{\rm s}}{3 a^{11}_{dd}}+\frac{\sqrt{a^{\alpha\alpha}_{\rm dd} a^{\beta\beta}_{\rm dd}}}{3a^{11}_{\rm dd}}(3\cos^2{\theta}-1)$, $\delta= u_{11} \rho_1 -u_{22} \rho_2$. For simplicity, we assume that the atomic masses of the two components are equal (i.e., $m_1=m_2=m$), which is justified for the typical Dy-Dy \cite{Santos:PRL:2021} as well as reasonable for Dy-Er \cite{Blakie:PRL:2021} mixtures, and the above equations have been nondimensionalized through scaling spatial coordinates and time by $\emph{l}=12\pi a^{11}_{\rm dd}$ and $m\emph{l}^2/\hbar$, respectively. Hereafter, our discussion will focus on the Dy-Er mixture, i.e., $a^{11}_{\rm dd}=132 a_0$ and $a^{22}_{\rm dd}=65.5a_0$ with $a_0$ being the Bohr radius.

\section{Results}

In this section we present that both in single- as well as in two-component dipolar BECs we can find metastable states that can even feature two length scales, despite the fact that there is only one roton minimum in the dispersion relation. For that matter, in the first subsection~\ref{susection:A} we illustrate in a single component BEC that new states with multiple length scales can be thought of as a certain superposition of other metastable states. 
Then, in~\ref{susection:B} we show that similar arguments also apply for two-component systems. 

\subsection{Single-Component BECs}\label{susection:A}

Let us start with the single component system and present the groundstate phase-diagram that has already been presented for a finite density, i.e. trapped in all three spatial directions,  in~\cite{Yongchang:PRA:2021,Pfau:PRR:2021} and in the thermodynamic limit including the stripe phase in~\cite{Yongchang:ATOMS:2022,Blakie:arxiv:2023}.

\begin{figure}
    \centering
    \includegraphics[width=\columnwidth]{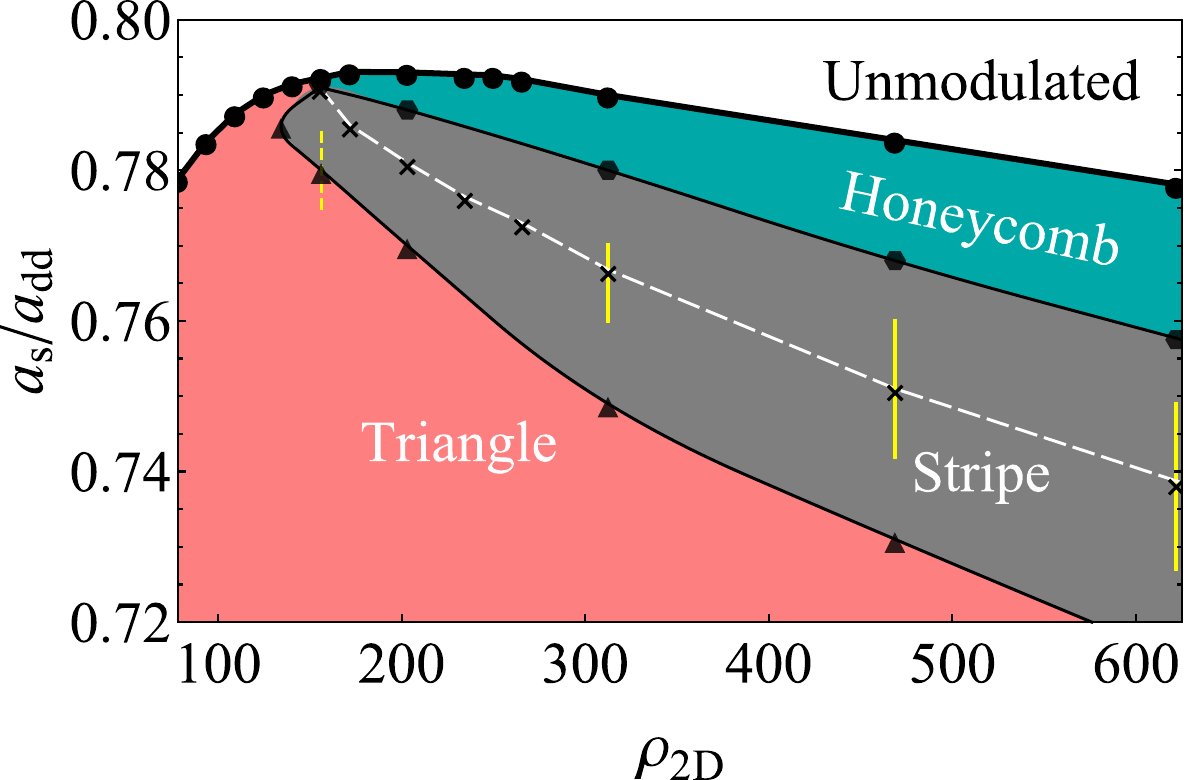}
    \caption{Groundstate phase diagram for a single component dipolar BEC. 
    The white dashed line separates the domain where patterns with stripe-symmetry are the groundstate into two regions, one where the (metastable) honeycomb lattice and one where the (metastable) hexagonal or triangular lattice are energetically preferred with respect to each other. 
    The white dashed line indicates where the latter become  energetically degenerate. The markers indicate numerically obtained points. 
    The newly added domain indicated by the yellow lines corresponds to regions where a surprising metastable ring state is supported [cf.~\reffig{fig:energies}(g)]. 
 }
    \label{fig:gs}
\end{figure}

The groundstate phase-diagram is shown in~\reffig{fig:gs}. 
We find that all phases emerge from the point where the superfluid-supersolid phase-transition becomes second-order and coexistence terminates. 
The region where the stripe-phase is the ground state can be separated into two areas, depending on whether the honeycomb (down-hexagons) or the triangular (hexagonal) states are energetically favorable with respect to each other. This transition is indicated by the white dashed line in~\reffig{fig:gs}.
Whereas in this area both the honeycomb as well as the triangular state are metastable, they feature remarkable stability and robustness. 
Therefore, it appears reasonable to ask whether this system robustly supports also more complex metastable states, such as states with multiple length-scales, two examples of which we present in the following. 

A natural approach for finding metastable states that feature more involved geometries is to inspect regions in the phase-diagram~\reffig{fig:gs} close to where the meta-stable states (i.e. honeycomb/triangular) feature equal energy, that is close to the earlier mentioned dashed line in~\reffig{fig:gs}. That is due to the fact that at these points the system does not favour either of them and, therefore, one might be tempted to expect that in this region they can be admixed in some way. This will be used to explain the emergence of surprising metastable phases in the next subsection. 

\subsubsection{Rings and ring-droplets}

\begin{figure}[!h]
    \centering
    \includegraphics[width=\columnwidth]{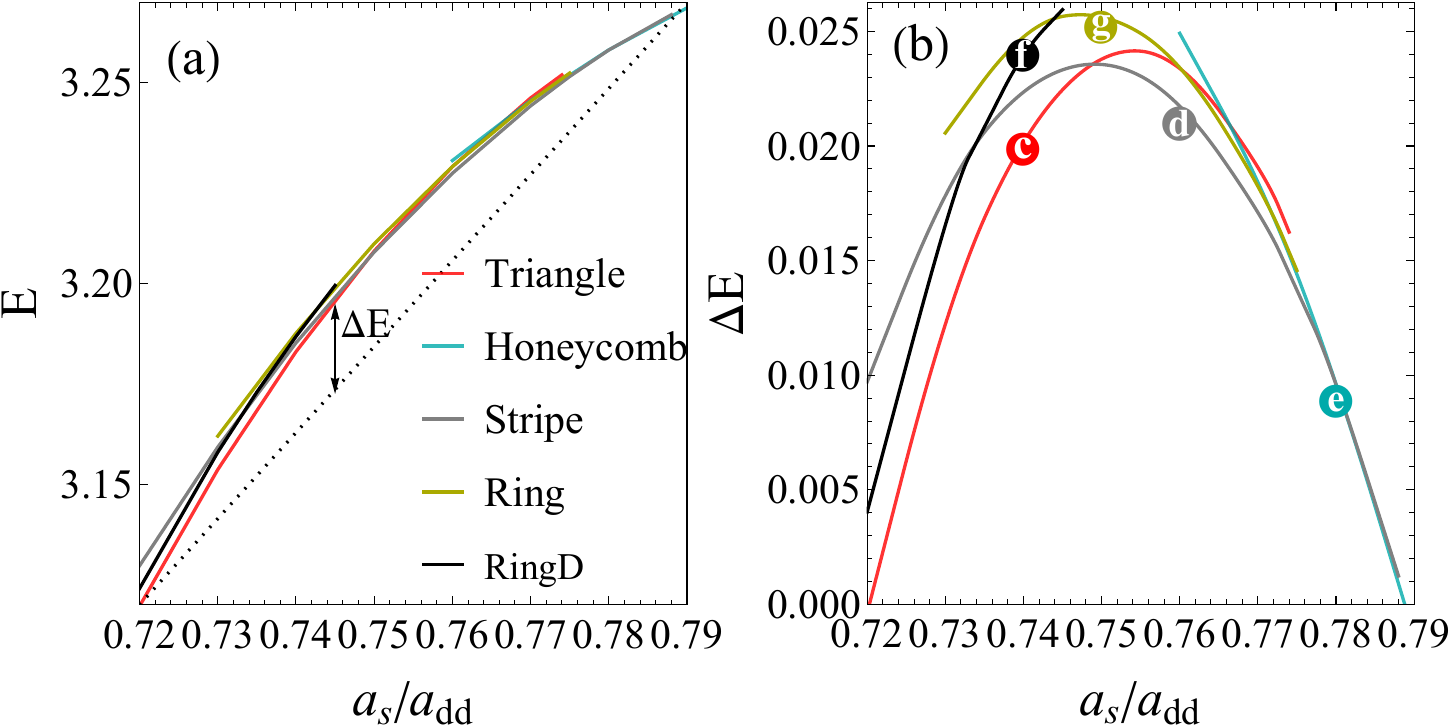}\\
    \includegraphics[width=\columnwidth]{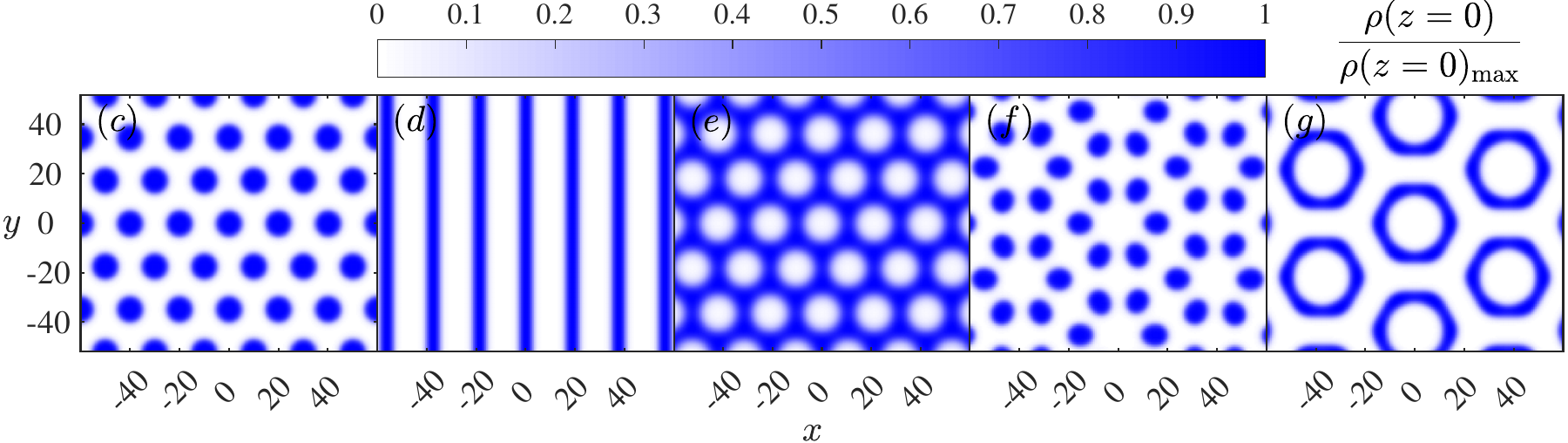}
    \caption{(a) depicts the energy landscape of the ground as well as metastable states at the density $\rho_{\rm 2D}=312.5$. The energy differences of each state with respect to the dashed line in (a) is plotted in (b) to represent the energies of different states more clearly. The density profiles at $z=0$ of the triangular, stripe, honeycomb, ring-droplet, and ring states are shown in the subplots (c-g), respectively. 
    Here the stripe state presented in (d), the ring state shown in (g) and the honeycomb state exhibited in (e) correspond to the phases at $a_{\rm s}/a_{\rm dd}=0.76$, $0.75$ and $0.78$, respectively, while the remaining states are at $a_{\rm s}/a_{\rm dd}=0.74$.
 }
    \label{fig:energies}
\end{figure}

Let us come back to the numerical results shown in~\reffig{fig:gs} and discuss the 
yellow region of~\reffig{fig:gs}, where isolated ring-like densities emerge [cf.~\reffig{fig:energies}(g)]. 
It appears that the yellow region shrinks upon approaching the second-order point, but does not converge to the second-order point. 
This can be expected, as only groundstates can converge to the second-order point, whereas supporting meta-stable states requires sufficiently large amplitude  modulations, as the latter permit the formation of a sufficiently deep local energy minima. 
Therefore, the numerical results matches the qualitative expectation: Initially the ``center-of-mass" of the ring-region appears to nicely follow the dashed transition-line upon decreasing density, yet at a certain point the domain has to stay sufficiently far away from the second-order point and deviates from that trend. 

Whereas the ring state is metastable with respect to the stripe phase, it features a lower energy than both the triangular as well as the honeycomb density distribution for the solid yellow lines. Upon deviating from the dashed white line, the ring state becomes energetically unfavourable compared to both triangular and honeycomb states, yet continues to exist. To distinguish that case we draw a dashed yellow line rather than a solid yellow line in~\reffig{fig:gs} closest to the second-order point.  

After this mostly qualitative discussion, let us now inspect the energies of the states involved in more detail. The energy per particle $E$ is given by the functional 
\begin{align}
 E&=\int \left(\frac{|\nabla\psi|^2}{2} + \frac{\omega^2_z z^2}{2}|\psi|^2+\frac{2}{5}\gamma N^{\frac{3}{2}}|\psi|^{5}\right){\rm d}{\bf r}+E_{\rm I}\\ 
E_{\rm I}&=\frac{N}{2}\!\!\int\!\left(\frac{a_{\rm s}}{3a_{\rm dd}}|\psi({\bf r})|^4+\!\int\!|\psi({\bf r})|^2 V({\bf r}-{\bf r}^\prime)|\psi({\bf r^\prime})|^2{\rm d}{\bf r^\prime} \right) {\rm d}{\bf r}\label{eq:Epot}
\end{align}
where $N$ is the total particle number and the wave function $\psi({\bf r})$ has been normalized to 1. 

The situation is plotted in~\reffig{fig:energies} for a fixed density $\rho_{\rm 2D}=312.5$.~\reffig{fig:energies}(a) shows the comparison of the energies of 
different states supported by the system in the region where the striped density distribution is the ground state as function of the $a_{\rm s}/a_{\rm dd}$. To add further visual clarity,~\reffig{fig:energies}(b) shows the energy difference between the respective states to the straight dashed line as depicted in~\reffig{fig:energies}(a). 

We find that for small values of $a_{\rm s}/a_{\rm dd}$ the ground state is given by a triangular distribution of density droplets~\reffig{fig:energies}(c). 
Upon increasing $a_{\rm s}/a_{\rm dd}$ sufficiently the groundstate corresponds to a stripe phase shown in~\reffig{fig:energies}(d) and finally, for large values of $a_{\rm s}/a_{\rm dd}$ is a honeycomb as depicted in~\reffig{fig:energies}(e). 

Let us now discuss the metastable states of~\reffig{fig:energies}(b) shown in~\reffig{fig:energies}(f,g) and start with the domain where the triangular or hexagonal lattice is the ground state. One might be tempted to expect that the stripe state ought to be the first metastable state, as it becomes the groundstate for larger $a_{\rm s}/a_{\rm dd}$. However, for small $a_{\rm s}/a_{\rm dd}$ the first metastable state we find corresponds to a droplet lattice that features two length scales [see~\reffig{fig:energies}(f)]. Upon further increasing $a_{\rm s}/a_{\rm dd}$ we find that the stripe phase becomes the first metastable state before becoming the groundstate. In the region where the stripe phase is the groundstate we can again distuinguish two regions, one where the triangular lattice is the first metastable state (small $a_{\rm s}/a_{\rm dd}$) and one that emerges upon increasing $a_{\rm s}/a_{\rm dd}$. This state is curious, as it features a density distribution that is ring-like [see~\reffig{fig:energies}(g)] and is rarely seen in other pattern-forming systems. The state shown in~\reffig{fig:energies}(f) resembles the state~\reffig{fig:energies}(g), as it carries the same underlying symmetry apart from an additional azimuthal modulation along the ring. 

Let us now return to the discussion of~\reffig{fig:gs}, but from the perspective of the length scales involved rather than the earlier presented energetic arguments. The deviation of the ``center-of-mass" of the ring-state region from the white dashed line can be understood as follows. At the second-order point co-existence of all phases with a single, fixed wave-vector converge. This wave vector $k$ can be found by the Bogoliubov excitation of the unmodulated state. In order to obtain a density distribution that features multiple length scales, the dynamics needs to be ``sufficiently nonlinear" in the modulation amplitude, i.e., sufficiently far away from the second-order point at which there can only be one length scale $\lambda=2\pi/k$. It is the nonlinearity that gives rise to a second length scale. This is again consistent with the fact that these ring states typically feature a large contrast and we were not able to find ring states with small modulation amplitude. 

Rings on a triangular lattice are dramatically different from the earlier mentioned more common patterns, as they can be thought of as a mix of both triangular and honeycomb lattice. To understand how they emerge and to furthermore identify the two earlier mentioned length-scales, let us  superpose two states such that metastable ring states close to the white dashed line in~\reffig{fig:gs} in the following fashion: 
\begin{equation}
    \rho^{\rm T, H}({\bf r}_{\perp},z)=\rho_0(z)\left( 1 \pm A \sum_{j=1}^3  \cos ({\bf k}_{j}\cdot{\bf r}_{\perp}+\varphi_j) \right).
    \label{eq:pert_TF}
\end{equation} 
Here, we assume $0<A\ll1$ as an ansatz for a weakly modulated condensate with triangular symmetry $\rho^{\rm T}$ or honeycomb symmetry $\rho^{\rm H}$, respectively. 
$A$ denotes the small amplitude of the density modulation. The three wave vectors form an equilateral triangle in the transverse plane with ${\bf k}_{1}+{\bf k}_{2}+{\bf k}_{3}=0$ and $|{\bf k}_{j}|=k$. 

To see how we can obtain a ring state with these states, let us superpose a triangular droplet lattice $\rho^{\rm T}$ with a honeycomb lattice $\rho^{\rm H}$ in the following manner: 
\begin{equation}
    \rho^{\rm R} = \mathcal{T}\mathcal{R}(\phi) \left[\rho^{\rm H}({\bf r}_{\perp},z)\right] - \rho^{\rm T}({\bf r}_{\perp},z)
\end{equation}

\begin{figure}[!h]
    \centering
    \includegraphics[width=\columnwidth]{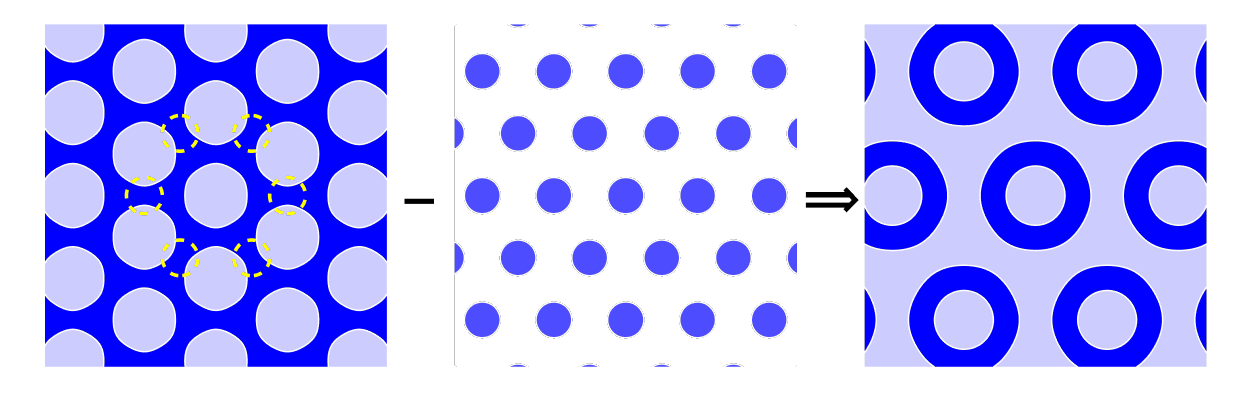}
    \caption{(a) shows a honeycomb lattice $\rho^{\rm H}$ that has been shifted to locate its minimum on top of the maximum of the triangular state and rotated by $\pi/6$, (b) a triangular lattice $\rho^{\rm T}$, (c) the superposition of the two lattices by subtracting (b) from (a).}
    \label{fig:superposition}
\end{figure}

Here, $\mathcal{R}(\phi)$ denotes a rotation perpendicular to the polarization direction by an angle of $\phi$ and $\mathcal{T}$ represents a translational shift operation. The idea of this ansatz is, basically, that the ring state looks similar to a honeycomb lattice with certain connections [see the yellow dashed circles in \reffig{fig:superposition}] being removed. To remove these connections as well as the background, we subtracted a triangular lattice. To further clarify the situation, the superposition process is shown in~\reffig{fig:superposition}. 

\begin{figure}[!h]
    \centering
    \includegraphics[width=\columnwidth]{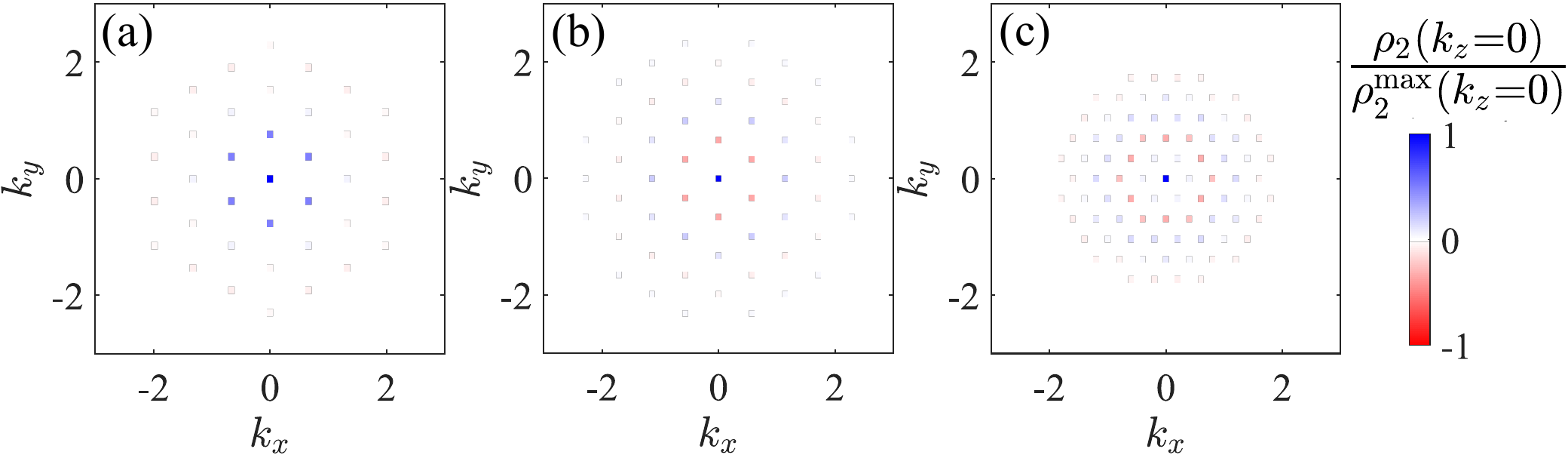}
    \caption{Fourier transforms of the (a) triangular, (b) honeycomb, and (c) ring states. This shows that ring state emerges as a mixture between a honeycomb lattice with a triangular lattice.}
    \label{fig:Fourier}
\end{figure}

This ansatz appears to capture the essence of the ring state, and we can directly read off that there are two sets of wavevectors $\bf k$: ${\bf k}_n=\frac{\sqrt{3}\pi}{a}(\cos(2\pi/n),\ \sin(2\pi/n))$  and ${\bf k}^\prime_n=\frac{2\pi}{a}(\cos(2\pi/n+\pi/6),\ \sin(2\pi/n+\pi/6))$ with $n=1\dots 6$ [also see \reffig{fig:Fourier}(c)]. Here, $a$ denotes the lattice constant. 

Surprisingly, even upon employing variational analysis, we find that such ring states can indeed be energetically preferable compared to both honeycomb as well as triangular lattice state, however, the stripe phase remains the ground state. 
To make sure that this result is not an artificial feature due to the variational ansatz, we show numerical results to confirm both existence as well as stability of such ring states. 

To validate whether this ansatz is consistent with the states we found numerically earlier in~\reffig{fig:energies}, we also show the Fourier transform of the ring state as well as the triangular and honeycomb states in~\reffig{fig:Fourier}. Evidently, the symmetry is correct and the Fourier transform shows what has been expected from this linear expansion~\footnote{It is useful to note that the Fourier transform of rotated state is the same as the rotated Fourier transform of the state.}. 

Interestingly, this state resembles the so-called fairy circles that can be found in waterscarce areas~\cite{Juergens:Science:2013,Getzin:PNAS:2016,Tarnita:Nature:2017,Getzin:Ecosphere:2019}, as it shares the property of multiple inherent length-scales (ring-size and ring-to-ring distance). Moreover, we would like to point out this ring state solely results from the strong nonlinear effect of dipolar condensates. This is in sharp contrast to similar phenomena observed in spin-orbit coupled BECs as well as unbalanced binary atomic mixtures, where the ring-like lattices are led by either gauge fields or external trapping with a ring geometry~\cite{VortexLatt:PRL,DipolarVortex:Nat.Phys.,Shi:PRA.104.013511,Berenstein:PRL:2003}.

\subsection{Two-component BECs}\label{susection:B} 

It is not priori clear whether our findings in the single-component case can be simply transferred to two-component systems. The biggest difference is the possibility to form miscible and immiscible or layered structures, and due to the cross-component interaction these layers interact. This interaction can be thought of as either being inhibiting or catalysing the formation of metastable states with multiple length scales. 

As a starting point to explore this more complicated system it is reasonable to first analytically write down an expression that captures the immiscibility in a two component system neglecting the modulations. In other words, we first consider unmodulated states and only focus on the new feature of immiscibility in this simplified system. After that we aim at transcribing the findings from single-component to the two-component case. 

\subsubsection{Miscibility of two-component unmodulated states}\label{sec:mis}

In this section we seek to find analytical approximations for the unmodulated density distributions including that captures their miscibility. 
As mentioned before, two-component systems can have different degrees of miscibility depending on the parameter regime~\cite{Cornell:PRL:1998,Wiemann:PRL:2008,Santos:PRL:2021,Blakie:PRL:2021,Lee:PRA:2021:Mixture}. 

\begin{figure}[h]
    \centering
    \includegraphics[width=\columnwidth]{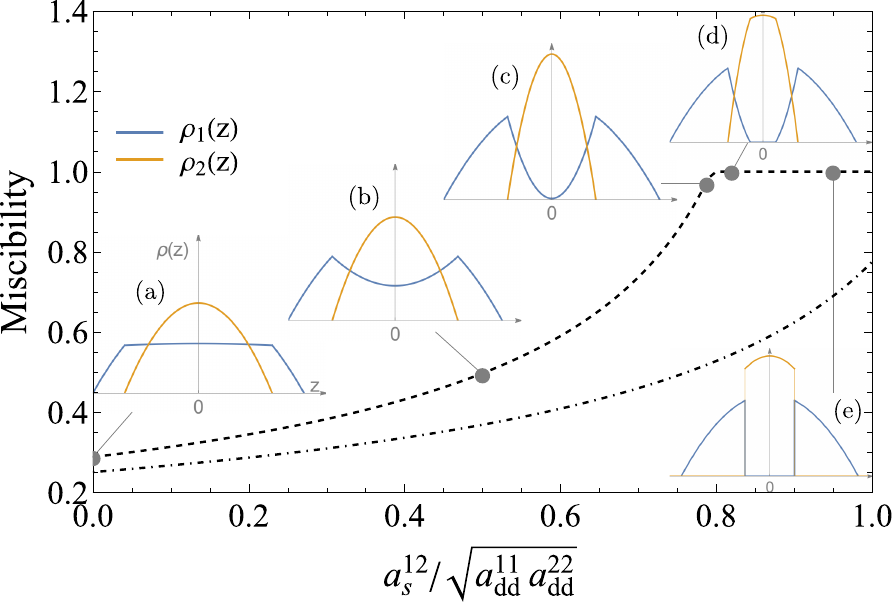}
    \caption{
    The miscibility $\frac{\rho_2(z=0)-\rho_1(z=0)}{\rho_2(z=0)+\rho_1(z=0)}$ is shown as function of $a^{12}_{\rm s}/\sqrt{a^{11}_{\rm dd}a^{22}_{\rm dd}}$. The insets show different profiles. 
    We fixed the density to $\rho^{\rm 2D}_1=\rho^{\rm 2D}_2=625$ and intra-component interactions to $a^{11}_{\rm s}/a^{11}_{\rm dd}=a^{22}_{\rm s}/a^{22}_{\rm dd}$. The dashed and the dotted-dashed lines correspond to $a^{11}_{\rm s}/a^{11}_{\rm dd}=0.9$ and $1.2$, respectively [cf. the horizontal dashed lines in~\reffig{fig:PhaseMixture}]. 
    }
    \label{fig:miscibility}
\end{figure}

Here, we consider an unmodulated state for both components and approximate the density distribution (i.e., $\rho_1$ and $\rho_2$) of that state with a Thomas-Fermi profile in the trapping direction $z$. After a bit of algebra, that allows to find 
\begin{widetext}
\begin{align} 
      \rho_1(z)&=
   \begin{cases}
   \frac{3a_{\rm dd}^{11}\left[\left(\mu_1-\omega_z^2z^2\right)\left(a_{\rm s}^{22}+2a_{\rm dd}^{22}\right)
                                -\left(\mu_2-\omega_z^2z^2\right)\left(a_{\rm s}^{12}+2\sqrt{a_{\rm dd}^{11}a_{\rm dd}^{22}}\right)\right]}
                                {                {(a_{\rm s}^{11}+2a_{\rm dd}^{11})(a_{\rm s}^{22}+2a_{\rm dd}^{22})-(a_{\rm s}^{12}+2\sqrt{a_{\rm dd}^{11}a_{\rm dd}^{22}})^2}} &\mbox{if } \rho_2(z)>0\\
   \frac{3a_{\rm dd}^{11}\left(\mu_1-\omega_z^2z^2\right)}
        {a_{\rm s}^{11}+2a_{\rm dd}^{11}}  & \mbox{if } \rho_2(z)=0 \\
\end{cases} \nonumber \\
  \rho_2(z)&=\frac{3a_{\rm dd}^{11}\left[\left(\mu_2-\omega_z^2z^2\right)\left(a_{\rm s}^{11}+2a_{\rm dd}^{11}\right)
                                -\left(\mu_1-\omega_z^2z^2\right)\left(a_{\rm s}^{12}+2\sqrt{a_{\rm dd}^{11}a_{\rm dd}^{22}}\right)\right]}
                {(a_{\rm s}^{11}+2a_{\rm dd}^{11})(a_{\rm s}^{22}+2a_{\rm dd}^{22})-(a_{\rm s}^{12}+2\sqrt{a_{\rm dd}^{11}a_{\rm dd}^{22}})^2} 
\label{eq:unmodulated_state}
\end{align}
\end{widetext}
with $\mu_1=\frac{1}{2}\omega^2_z\sigma^2_{z1}$, $\mu_2=\omega^2_z(\sigma^2_{z1}-\sigma^2_{z2})\frac{a^{12}_{\rm s}+2\sqrt{a^{11}_{\rm dd}a^{22}_{\rm dd}}}{a^{11}_{\rm s}+2a^{11}_{\rm dd}}+\frac{1}{2}\omega^2_z\sigma^2_{z2}$, $\sigma_{z1}=\left[\frac{\rho_{\rm 2D}(a^{11}_{\rm s}+a^{12}_{\rm s}+2a^{11}_{\rm dd}+2\sqrt{a^{11}_{\rm dd} a^{22}_{\rm dd}})}{2a^{11}_{\rm dd} \omega^2_z} \right]^{1/3}$, $\sigma_{z2}=\left[ \frac{\rho_{\rm 2D}(a^{11}_{\rm s} a^{22}_{\rm s} +2a^{11}_{\rm s} a^{22}_{\rm dd}+2 a^{22}_{\rm s} a^{11}_{\rm dd}-a^{12}_{\rm s} a^{12}_{\rm s} -4a^{12}_{\rm s} \sqrt{a^{11}_{\rm dd} a^{22}_{\rm dd}})}{2a^{11}_{\rm dd} \omega^2_z (a^{11}_{\rm s}-a^{12}_{\rm s}+2a^{11}_{\rm dd}-2\sqrt{a^{11}_{\rm dd} a^{22}_{\rm dd}})} \right]^{1/3}$. Here, we assumed the same average 2D density $\rho^{\rm 2D}_1=\rho^{\rm 2D}_2=\rho_{\rm 2D}$ for the two components.  $\rho_2$ is the component that remains localized around $z=0$ and $\rho_1$ gets pushed out in the immiscible regime. That is reflected by the fact that $\rho_1$ is a piece-wisely defined function. \refeq{eq:unmodulated_state} motivates to introduce the parameter
\begin{equation}
    \eta=\frac{a_{\rm s}^{12}+2\sqrt{a_{\rm dd}^{11}a_{\rm dd}^{22}}}{a_{\rm s}^{22}+2a_{\rm dd}^{22}}
    \label{eq:gamma}
\end{equation}
For $\eta=1$ represents the point where the transition from the miscible to the immiscible regime occurs. In the miscible regime ($\eta<1$), both $\rho_1$ and $\rho_2$ reach maximum value at $z=0$ as can be seen from~\reffig{fig:miscibility}(a). In contrast, for $\eta>1$ the density of $\rho_1$ becomes a parabola close to $z=0$ [see \refeq{eq:unmodulated_state} and \reffig{fig:miscibility}(b,c)]. Hence, $\rho_1$ no longer has its maximum at $z=0$, but it acquires two maxima at the point where $\rho_2(z)$ becomes zero, while $\rho_2$ retains the normal Thomas-Fermi profile. In contrast, both $\rho_1$ and $\rho_2$ deviates from the normal Thomas-Fermi distribution when $\eta>\mu_1/\mu_2$, where the strong contact inter-species repulsion eventually fully separates the two components as shown in \reffig{fig:miscibility}(d,e)]. In that case, $\rho_1$ vanishes in a finite region around $z=0$. In other words the binary condensates enter the immiscible regime. 

In the strong immiscible regime the density profiles of the two species can be expressed as below
\begin{widetext}
\begin{align}
	\rho_1(z)&=
	\begin{cases}
		0 & \mbox{if } |z|\leq \sigma_{z3} \\
		\frac{3a_{\rm dd}^{11}\left(\mu_1-\omega_z^2z^2\right)}
		{a_{\rm s}^{11}+2a_{\rm dd}^{11}}  & \mbox{if } \sigma_{z2}\leq |z|\leq \sigma_{z1} \\
		\frac{3a_{\rm dd}^{11}\left[\left(\mu_1-\omega_z^2z^2\right)\left(a_{\rm s}^{22}+2a_{\rm dd}^{22}\right)
			-\left(\mu_2-\omega_z^2z^2\right)\left(a_{\rm s}^{12}+2\sqrt{a_{\rm dd}^{11}a_{\rm dd}^{22}}\right)\right]}
		{                {(a_{\rm s}^{11}+2a_{\rm dd}^{11})(a_{\rm s}^{22}+2a_{\rm dd}^{22})-(a_{\rm s}^{12}+2\sqrt{a_{\rm dd}^{11}a_{\rm dd}^{22}})^2}} &\mbox{if } \sigma_{z3}\leq |z|\leq \sigma_{z2}
	\end{cases} \nonumber \\
	\rho_2(z)&=
	\begin{cases}
	    \frac{3a_{\rm dd}^{11}\left(\mu_2-\omega_z^2z^2\right)}
		{a_{\rm s}^{22}+2a_{\rm dd}^{22}} & \mbox{if } |z|\leq \sigma_{z3} \\
		\frac{3a_{\rm dd}^{11}\left[\left(\mu_2-\omega_z^2z^2\right)\left(a_{\rm s}^{11}+2a_{\rm dd}^{11}\right)
		-\left(\mu_1-\omega_z^2z^2\right)\left(a_{\rm s}^{12}+2\sqrt{a_{\rm dd}^{11}a_{\rm dd}^{22}}\right)\right]}
	{(a_{\rm s}^{11}+2a_{\rm dd}^{11})(a_{\rm s}^{22}+2a_{\rm dd}^{22})-(a_{\rm s}^{12}+2\sqrt{a_{\rm dd}^{11}a_{\rm dd}^{22}})^2} &\mbox{if } \sigma_{z3}\leq |z|\leq \sigma_{z2}
	\end{cases}, 
	\label{eq:unmodulated_state1}
\end{align}
\end{widetext}
where the chemical potentials $\mu_{1,2}$ take the same form as in the former case. The critical points $\sigma_{z1,z2,z3}$ can be determined by the constraint $\int \rho_{1,2}(z)dz=\rho_{\rm 2D}$. For the case of intermediate cross interaction [cf. Fig.~\ref{fig:PhaseMixture}(d)], there remains overlap of the two species in the region $\sigma_{z3}\leq|z| \leq \sigma_{z2}$. However, if the inter-species contact interaction exceeds the critical strength $(a^{12}_{\rm s})_c=\sqrt{(a^{11}_{\rm s}+2a^{11}_{\rm dd})(a^{22}_{\rm s}+2a^{22}_{\rm dd})}-2\sqrt{a^{11}_{\rm dd}a^{22}_{\rm dd}}$, the overlap between the two components disappears completely [see Fig.~\ref{fig:PhaseMixture}(e)] and it enters the completely immiscible regime.

\subsubsection{Excitation spectrum of two-component unmodulated states}\label{sec:exc_spec}

For convenience we report here an (approximated) excitation spectrum of two-component unmodulated states as their expressions are useful for the next subsection. The (approximated) excitation spectrum can be easily obtained  from linearisation of the governing equation of motion \refeq{eq:NLS}. For that matter, we use the expression obtained before~\refeq{eq:unmodulated_state} for $\rho_1(z),\rho_2(z)$ and integrate out the $z$-direction. With that we obtain the following expression that only depends on $k_{\perp}$ as follows:

\begin{align}
\omega^2=&\frac{k_{\perp}^2}{2}\bigg[
\frac{k_{\perp}^2}{2}+ \rho^{\rm 2D}_1 g_{11}
+ \rho^{\rm 2D}_2 g_{22}+ \rho^{\rm 2D}_1 v_{11}+ \rho^{\rm 2D}_2 v_{22}\nonumber\\
&- \bigg( \left(\rho^{\rm 2D}_1 g_{11}+\rho^{\rm 2D}_1 v_{11}-\rho^{\rm 2D}_2 g_{22}-\rho^{\rm 2D}_2 v_{22}\right)^2\nonumber \\ 
&+ 4 \rho^{\rm 2D}_1 \rho^{\rm 2D}_2 (g_{12}+v_{12})^2\bigg)^{\frac{1}{2}}\bigg].
\label{eq:ex_spec}
\end{align}
Here $\rho^{\rm 2D}_{\alpha}=\int \rho_{\alpha}(z){\rm d}z$ ($\alpha=1,\ 2$) denotes the average 2D density of the condensate in the plane perpendicular to the polarization direction and $v_{\alpha \beta}({\bf k}_\perp)=\frac{\sqrt{a^{\alpha \alpha}_{\rm dd} a^{\beta \beta}_{\rm dd}}}{a^{11}_{\rm dd} \rho^{\rm 2D}_{\alpha}\rho^{\rm 2D}_{\beta}}\int \mathcal{F}[\rho_{\alpha}(z)] \mathcal{F}[\rho_{\beta}(z)] \left(\frac{k^2_z}{k^2_\perp+ k^2_z}-\frac{1}{3} \right){\rm d}k_z$. $\mathcal{F}[\rho_{\alpha}]$ corresponds to the Fourier transform of $\rho_\alpha$ and $g_{\alpha \beta}=\frac{a^{\alpha\beta}_{\rm s}}{3 a^{11}_{\rm dd} \rho^{\rm 2D}_{\alpha}\rho^{\rm 2D}_{\beta} } \int \rho_{\alpha}(z) \rho_{\beta}(z) {\rm d}z $ represents the effective 2D contact interaction. To obtain the above simple analytical excitation spectrum, we have neglected the LHY correction terms. This is not at all justified and serves for a qualitative discussion only and is certainly quantitatively wrong. 

\begin{figure}[h]
    \centering
    \includegraphics[width=\columnwidth]{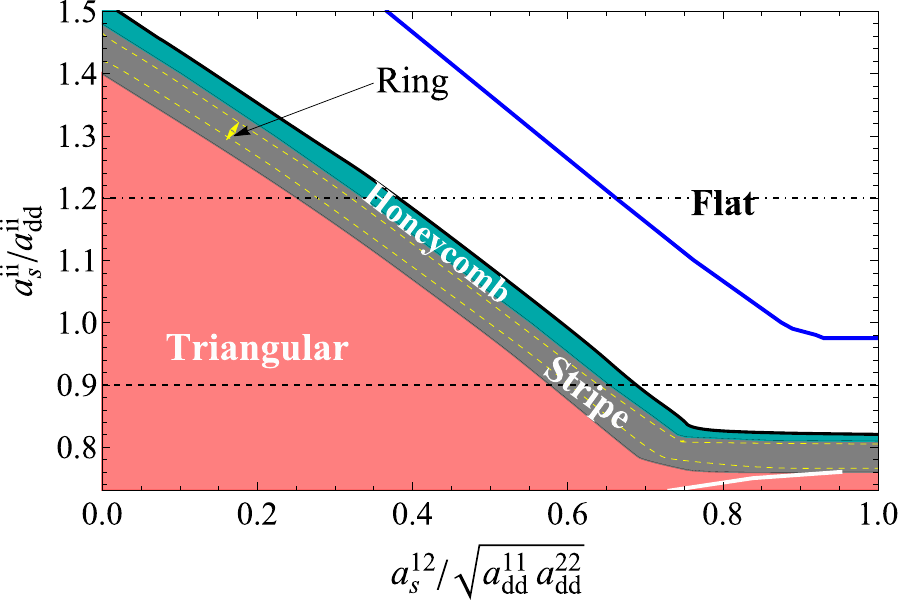}
    \caption{The phase diagram of two-component dipolar condensates with fixed density of $\rho^{\rm 2D}_1=\rho^{\rm 2D}_2=625$ and balanced intra-component interactions $a^{11}_{\rm s}/a^{11}_{\rm dd}=a^{22}_{\rm s}/a^{22}_{\rm dd}$. Again, the ring state remains metastable with respect to the stripe phase, however, in the region indicated by the dashed yellow line it is energetically preferred compared to the triangular and honeycomb states. The overall trend of all lines can be estimated via the approximated excitation spectrum~\refeq{eq:ex_spec}, which amounts to the blue solid line. 
    The two horizontal lines correspond to the values of $a^{ii}_{\rm s}/a^{ii}_{\rm dd}$ ($i=1,2$) for which the miscibility $\frac{\rho_2(z=0)-\rho_1(z=0)}{\rho_2(z=0)+\rho_1(z=0)}$ is shown in~\reffig{fig:miscibility}. We find that the ring state persists for a large range of miscibility. Furthermore, for the values at hand, it appears that the miscibility does not affect the existence regions of the states, as despite of a dramatic change in the former, the overall trend guessed from the (approximated) excitation spectrum remains. 
    The domain of triangular states is separated into two regions by the white solid line. Left to this line the groundstate is the usual triangular lattice, whereas right to this line the groundstate  features the triangular superlattice that will be discussed further in subsection~\ref{sec:triangular_superlattice}. 
    }
    \label{fig:PhaseMixture}
\end{figure}

The blue solid line in figure~\reffig{fig:PhaseMixture} shows the position where he roton minimum of the unmodulated state's excitation spectrum touches zero as function of the scattering length $a^{12}_{\rm s}$. 
Remarkably, this line displays distinct behaviors in the weak and strong cross contact interaction regime. In the case of small scattering length $a^{12}_{\rm s}$, the roton instability can be promoted either by the inter-component interaction $a^{12}_{\rm s}$ or by the intra-component interaction $a^{ii}_{\rm s}$ under a general constraint.

In contrast, in the strong cross interaction regime, this critical line no longer depends on $a^{12}_{\rm s}$ and is solely determined by the intra-component interaction $a^{ii}_{\rm dd}$. This can be understood as follows. As can be seen from \reffig{fig:miscibility}(e), there is no overlap between the distributions of the two components at large $a^{12}_{\rm s}$. Therefore, $g_{12}$ becomes zero and the excitation spectrum $\omega$ depends on the intra-component contact interaction only [see \refeq{eq:ex_spec}]. In contrast, in case of small $a^{12}_{\rm s}$, the two components are miscible and their overlap is determined by their cross interaction. This is why the excitation spectrum exhibits the two different scalings. 

The qualitative behavior is confirmed by numerical simulations [see the black solid line in \reffig{fig:PhaseMixture}]. 
The significant difference between the full numerical results (black line) and the approximated excitation-spectrum (blue line) is clearly visible, and it is clear that for quantitatively correct results the LHY correction has to be taken into account. Since the LHY correction acts like contact repulsive interactions which tends to stabilize the unmodulated state, the roton-instability critical line would be shifted towards smaller $a^{ij}_{\rm s}$ by quantum fluctuations as the numerical results show. 

\subsubsection{Rings and ring-droplets in two-component systems}

We already established that we can find ring states in the single component system, however, it is unclear whether we can find their analogue in two-component systems. 
That is due to the fact that the different layers interact and thereby might prevent the formation of
ring-states. 

\begin{figure}[h]
    \centering
    \includegraphics[width=\columnwidth]{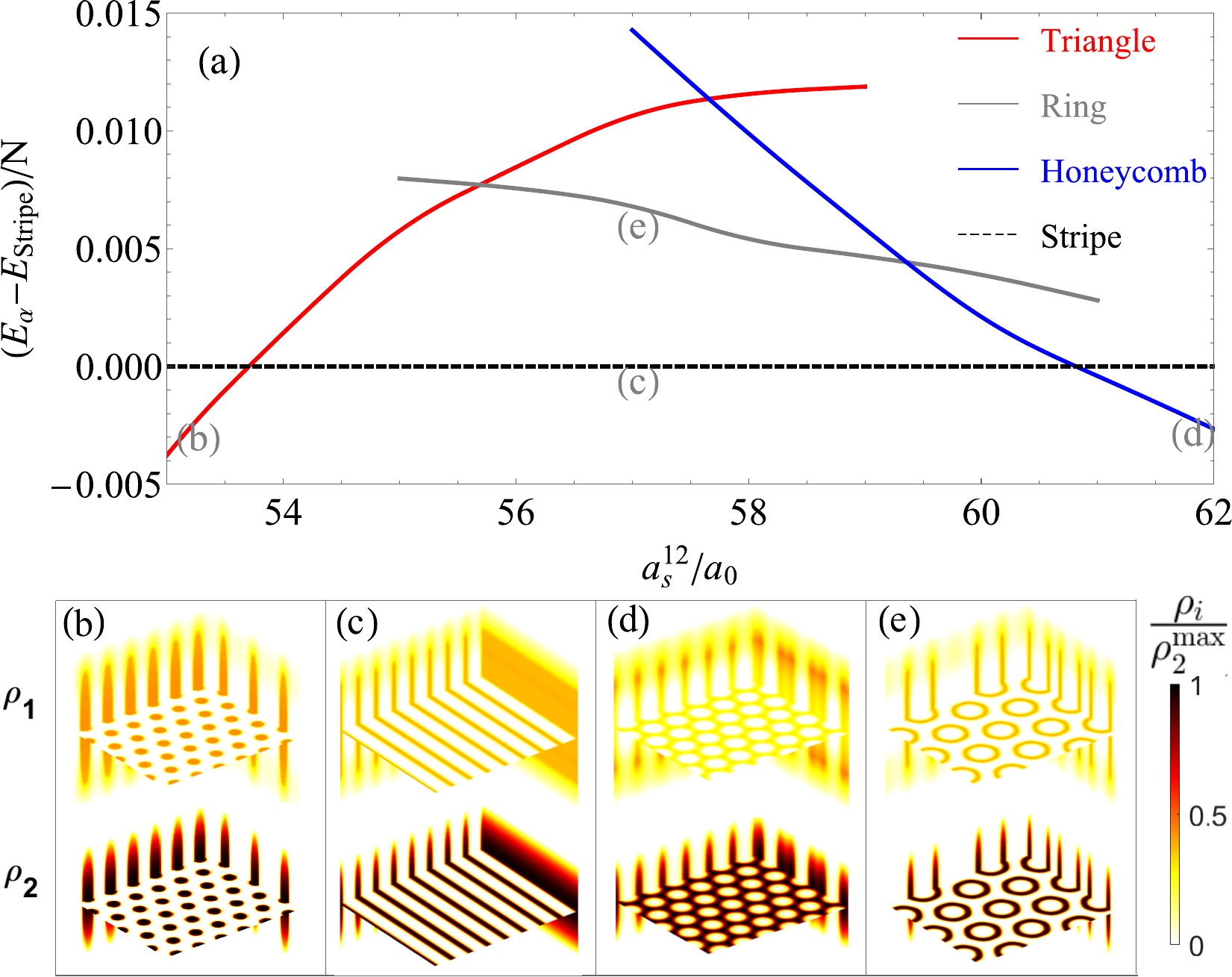}
    \caption{The energy for four different stable states is shown in (a) for $\frac{a^{11}_{\rm s}}{a^{11}_{\rm dd}}=\frac{a^{22}_{\rm s}}{a^{22}_{\rm dd}}=0.9$. The ground state for most of the considered region remains a stripe phase. In analogy to the the single component case shown in~\reffig{fig:gs}, we find that in a two-component BEC ring states are possible as well. 
    As in the single-component BEC, these ring states occur in a region close to where honeycomb and triangular lattices become energetically comparable. 
    }
    \label{fig:states}
\end{figure}

To explore that, consider~\reffig{fig:states}(a), which displays energy differences to the energy of the stripe phase. The supplementary figures (b,c,d,e) display the densities of the different states involved.  We find again that close to where the triangular and honeycomb state become energetically degenerate and the striped state is the ground state, a metastable ring state emerges. Close to the aforementioned energy degeneration, this ring state features a lower energy than both the triangular state as well as the honeycomb state, akin to what we already found in the single-component case. 

Now that we established that ring states are also possible in dipolar BEC mixtures, let us address whether their existence is promoted or suppressed when tuning the miscibility. As we have seen before in~\refeq{eq:gamma}, one can alter the miscibility by tuning the value of $a^{12}_{\rm s}$. 
The result is provided in~\reffig{fig:PhaseMixture}, where we show the phase diagram for a fixed density and varying cross-contact interaction and contact interaction in the upper panel (a). Here, we focus on the case of balanced intra-component interactions, i.e., $a^{11}_{\rm s}/a^{11}_{\rm dd}=a^{22}_{\rm s}/a^{22}_{\rm dd}$.

We see that for this set of parameters the regions of the different phases are practically unaffected despite tuning the miscibility dramatically from 0.3 to 1 [see \reffig{fig:PhaseMixture}]. In other words, the ring state, rather than ceasing to exist, even retains its domain of existence. This shows that the ring states are robust, as the cross-interaction can be considered as a perturbation to the single-component physics that deforms the energy landscape significantly. 

In fact, the opposite is true for the displayed case: The cross-interaction with the other component actually stabilizes the ring state as compare to the single-component case beyond the otherwise critical line of $a_{\rm s}/a_{\rm dd}$ shown in~\reffig{fig:gs} of around $a_{\rm s}/a_{\rm dd}\approx 0.79$, where in the single component case all modulated states cease to exist in favor of the unmodulated state. 
Therefore, in a way one might say that these somewhat peculiar states actually emerge in a broad parameter region in both single- as well as two-component dipolar BECs. Moreover, the cross interactions in two-component systems can stabilize the existence of these exotic states. 

\subsubsection{Superfluid fraction of modulated states}

\begin{figure}[!htb]
	\centering
	\includegraphics[width=\columnwidth]{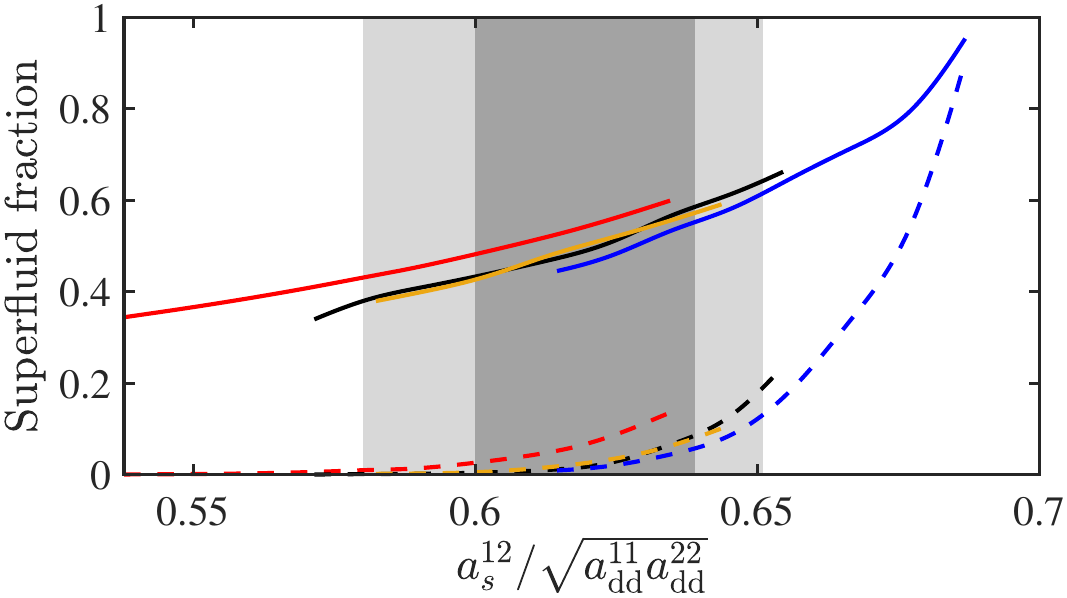}
	\caption{The superfluid fraction of different stable states vary with the interspecies contact interaction while the intraspecies interaction is fixed at  $a^{\rm 11}_{s}/a^{\rm 11}_{\rm dd}=a^{\rm 22}_{s}/a^{\rm 22}_{\rm dd}=0.9$, where the full (dashed) lines indicate the superfluid fraction of the first (second) component. Here the red, black, dark yellow, and blue lines represent the corresponding superfluid fraction $f^s_1$ ($f^s_2$) of the triangular, stripe, ring, and honeycomb states, respectively. The regime of the stripe ground state is denoted by the light gray color, while the dark gray shadow illustrates that the ring state is energetically favorable than the triangular and honeycomb states in this region.}
	\label{fig:SF}
\end{figure}


Thus far we did not yet consider the superfluid properties of the different patterns we found. We will do that now using Legget's bound~\cite{Leggett:PRL:1970,Leggett:JSP:1998}.
\begin{equation}\label{eq:SF}
	f^{\rm s}_\alpha = \min_\theta\left[\int\frac{L^2_x}{\int\rho_\alpha(\bar x,\bar y,z)^{-1}{\rm d} x}{\rm d} y{\rm d}z\right].
\end{equation}
where $\alpha=1,2$, $L_x$ is the size of the numerical box along $x$ direction, and we take the minimum with respect to all possible directions defined by the angle $\theta$ with $\bar x=x\cos\theta-y\sin\theta$ and $\bar y=x\sin\theta+y\cos\theta$~\cite{Yongchang:PRL:2019}.
This is a useful quantity to estimate the superfluid fraction of the condensate~\cite{Rica:PRB:2008,Santos:PRA:2022} and has recently been refined~\cite{Blakie:arxiv:2023b}. 
\reffig{fig:SF} presents the superfluid fraction of the triangular (red lines), stripe (black lines), ring (dark yellow lines) as well as honeycomb (blue lines) states in the vicinity of the critical region from modulated states to a flat state. Here, the interspecies interaction is fixed at $a^{\rm 11}_{s}/a^{\rm 11}_{\rm dd}=a^{\rm 22}_{s}/a^{\rm 22}_{\rm dd}=0.9$ (i.e., along the black dashed line in~\reffig{fig:PhaseMixture}). We note that the first component (solid  lines) possesses a large superfluid fraction compared to the superfluid fraction of the second component (dashed lines) which is practically negligible for small values of $a^{\rm 12}_{\rm s}$. 
However, for sufficiently large values of $a^{\rm 12}_{\rm s}$, such that the honeycomb can form in the second component (blue dashed line), also the second component is able to increase its superfluid fraction significantly due to the overlapping wavefunction, i.e. atoms can flow along the bridges of the honeycomb~\cite{Yongchang:PRL:2019}. 

To gain further insight into why the two components behave so distinctively different, let's consider the single-component case. For the second component this individual species has a much lower dipolar length (i.e. $a^{\rm 22}_{\rm dd}=65.5a_0$ vs. $a^{\rm 11}_{\rm dd}=132a_0$), it would require a much larger average 2D density to reach the second-order point where superfluidity is large and from where new phases with higher superfluid fraction emerge~\cite{Yongchang:PRL:2019}. 
Thus, for the case of balanced dipolar mixtures considered here, the second component remains almost insulating. Nevertheless, as stated before the situation changes when second component enters the honeycomb regime. 

\subsubsection{Deformed triangular states at strong cross interaction}\label{sec:triangular_superlattice}

As discussed in Sec.~\ref{sec:mis}, the two-component dipolar mixtures feature a transition from a miscible to an immiscible distribution that can be controlled by the cross contact interaction $a^{12}_s$. 
Thus far, the discussion of miscibility has been limited to polarisation direction as we considered states that are unmodulated in the transverse plane (perpendicular to the polarisation direction). 

What is missing is whether there is a parameter region of modulated states where we can see clear signatures of immiscibility in the transverse plane. 

As a proof-of-principle that such regions exist and to illustrate the effect of ``transverse immiscibility", we go into the region of deeply crystallized triangular states that appears for large values of $a^{12}_{\rm s}$ (cf.~\reffig{fig:PhaseMixture}, right to the white line). 

\begin{figure}[!htb]
	\centering
	\includegraphics[width=\columnwidth]{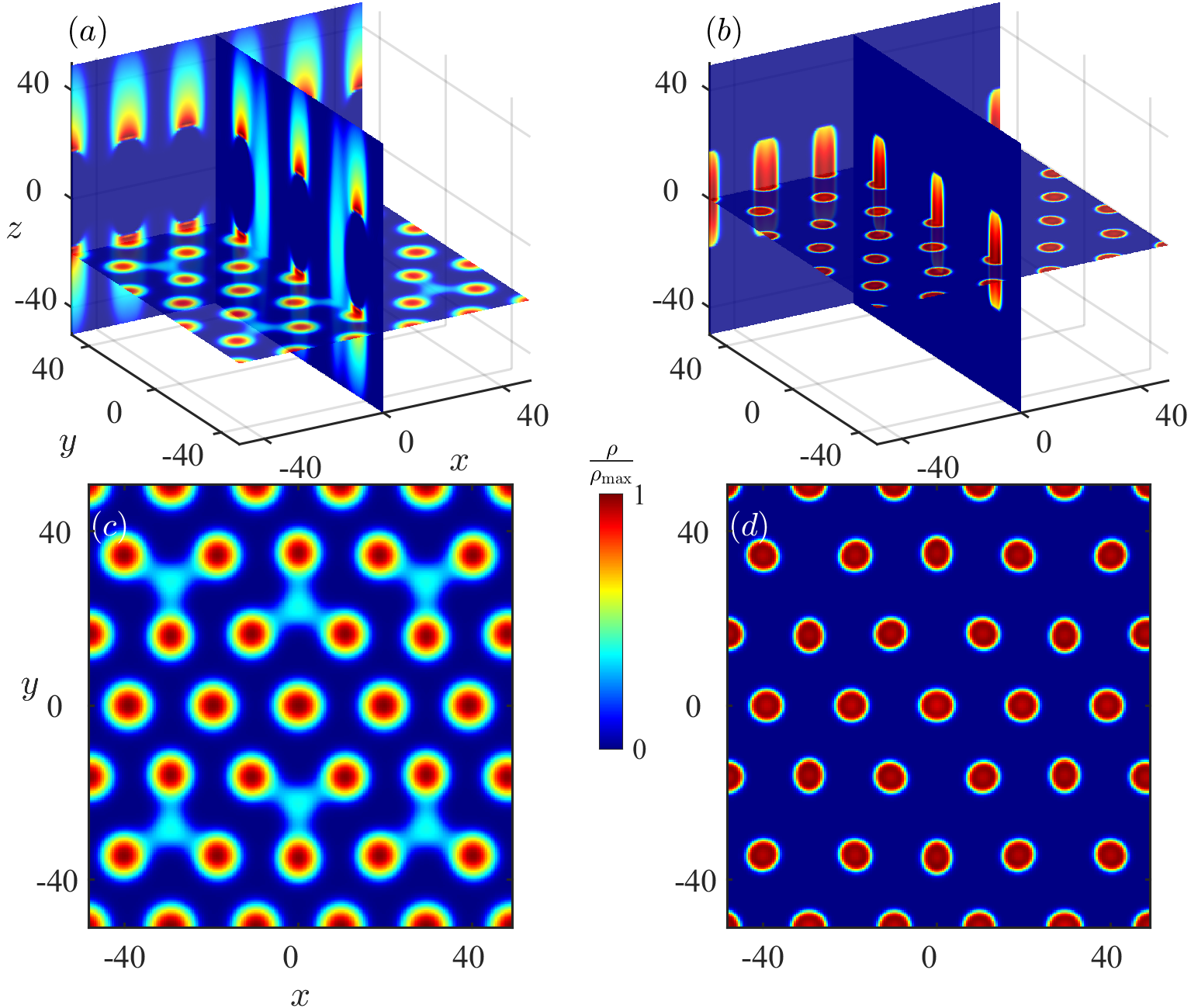}
	\caption{The density profiles of the deformed triangular states at strong inter-species contact interaction, i.e., the right side of the white line in Fig.~\ref{fig:PhaseMixture}(a), where the interactions are fixed at $a^{\rm ii}_{s}/a^{\rm ii}_{\rm dd}=0.74$ and $a^{12}_s/\sqrt{a^{11}_{\rm dd} a^{22}_{\rm dd}}=0.86$, while the density is the same as Fig.~\ref{fig:PhaseMixture}(a). Panels (a) and (b) present the 3D density profiles of $\rho_1$ and $\rho_2$, respectively, via the slice plots. To clearly see the geometry of such states, the 2D distributions $\rho_1(z=z_c)$ and $\rho_2(z=0)$ are shown in panels (c) and (d) as well. Here $z_c$ denotes the position where $\rho_1$ reaches its maximum in the $z$ direction.}
	\label{fig:TriangleS}
\end{figure}

Fig.~\ref{fig:TriangleS} shows an example of deformed triangular states at strong cross contact interaction $a^{12}_s/\sqrt{a^{11}_{\rm dd} a^{22}_{\rm dd}}=0.86$. From the 3D slice plot in the panels (a) and (b), one can discern that the first component is not fully separated from the second component in $z$ (not completely immiscible). Instead, new ``droplets" turn up at the center of three neighbouring regular droplets as shown in Fig.~\ref{fig:TriangleS}(a) and (c). Furthermore, these new ``droplets" spread across $z=0$ plane and  bridge between the upper and lower layers of the first component.
The reason is that the triangular state is composed of well separated droplets in the deep modulation regime. Upon increasing the cross contact interaction, the first component is pushed towards large $z$ in order to minimize the total energy by reducing the overlap between the two components. If $a^{12}_s$ is too large, part of the atoms prefer to stay around $z=0$ with low trapping potential energy, and eventually give rise to these new links between the two dominant layers. In this situation the second component can no longer maintain its regular triangular distribution and resorts to a deformed triangular profile as displayed in Fig.~\ref{fig:TriangleS}(d). As mentioned, the phase-boundary between regular and deformed triangular states is denoted by the white line in Fig.~\ref{fig:PhaseMixture}.  
In contrast, the fully immiscible honeycomb or flat states do not have links in between the outside layers, since the second component possesses a significant superfluid background which prevents the formation of such layer links due to the strong interspecies contact repulsion.

\section{Conclusions}

In this paper we established that dipolar BECs support unusual metastable robust states featuring multiple length scales. 
These states have the shape of a ring-like density distribution whose azimuthal density modulation can be modulated via tuning the scattering length. They appear in a domain where the stripe phase is the ground state and in a region around the line where the (metastable) triangular and honeycomb lattice become energetically degenerate. 

Moreover, these states can be stabilized in a much broader regime in the binary dipolar BECs where only unmodulated flat state exist in the single-component counterparts. In sharp contrast to other ring-like phenomena induced by gauge fields and ring-shape confinement in spin-orbit coupled or unbalanced quantum gas mixtures, the ring-lattice state reported here is solely induced by the strong nonlinear effects.

Although such ring states do not emerge as ground states, they underpin the variety of stable self-organized structures in long-range interacting systems, as already featured in, e.g.,~\cite{Lewenstein:PRL:2007,Lewenstein:PRA:2008,Pupillo:PRL:2010,Pfau:PRL:2012,Ferenc:PRL:2018}. Therefore, dipolar BECs represent a promising platform to explore metastable-state quantum phase transition as well \cite{Kosterlitz:JPC:1973,Ueda:PRA:2010}. 

In addition to the various metastable states emerging close to where the  triangular and honeycomb states become energetically degenerate, we also discovered a deformed triangular superlattice  groundstate in the deeply modulated and immiscible regime. 
Usually, immiscibility is discussed as a phenomenon occurring in the polarization direction (e.g.,~\cite{Santos:PRL:2021,Blakie:PRL:2021}) rather than the plane transverse to the polarization. 
In this case the emergence of the triangular superlattices with periodic density-bridges displayed in~\reffig{fig:TriangleS} is a clear new feature due to ``transverse immiscibility" of two-component BECs that is different to what has been discussed and has no analogue with a single component dipolar BEC. 

As an outlook, we think that further understanding of the topology of this phase-diagram can be found by studying the bifurcation diagram~\cite{Alina:2022}. Its dependency on temperature seems to represent an interesting endeavour as well~\cite{Oktel:PRA:2019, Ferlaino:PRL:2021, Pohl:2022}. Furthermore, in this work we restricted our consideration to a small subset of parameters, and extending that 
to e.g. unequal intra-component interactions and unequal masses to explore the rich spectrum of metastable states remains to be done. 

\section{Acknowledgement}

This work was supported by the National Nature Science Foundation of China (Grant No.: 12104359), National Key Research and Development Program of China (Grant No.: 2021YFA1401700), Shaanxi Academy of Fundamental Sciences (Mathematics, Physics) (Grant No.: 22JSY036), and the Danish National Research Foundation through the Center of Excellence ``CCQ" (Grant No.: DNRF156). Y.C.Z. acknowledges the support of Xi'an Jiaotong University through the ``Young Top Talents Support Plan" and Basic Research Funding as well as the High-performance Computing Platform of Xi'an Jiaotong University for the computing facilities. F.M. acknowledges the funding from the Ministerio de Economía y Competitividad (PID2021-128910NB-100).
\bibliography{bib.bib}

\end{document}